\documentclass[psfig]{mn2e}
\input epsf

\newcommand{\av}{$A_{V}$} 
\newcommand{\abj}{$A_{(B_J)}$} 
\newcommand{\alam}{$A_{\lambda}$}
\newcommand{\ebv}{ {\it E(B--V)}}
\newcommand{\ang}{$\rm \AA$}

\newcommand{\degree}{$^{\rm o}$}

\newcommand{\mic}{$\mu$m}

\newcommand{\ha}{H$\alpha$}

\newcommand{\bb}{\bibitem[]{bla}}

\usepackage{times}

\def\deg{\hbox{$^\circ$}}

\def\lesssim{\mathrel{\hbox{\rlap{\hbox{\lower4pt\hbox{$\sim$}}}\hbox{$<$}}}}
\def\gtrsim{\mathrel{\hbox{\rlap{\hbox{\lower4pt\hbox{$\sim$}}}\hbox{$>$}}}}

\def\ion#1#2{#1$\;${\small\rm\@Roman{#2}}\relax}

\newbox\grsign \setbox\grsign=\hbox{$>$} \newdimen\grdimen 
\grdimen=\ht\grsign
\newbox\simlessbox \newbox\simgreatbox
\setbox\simgreatbox=\hbox{\raise.5ex\hbox{$>$}\llap
      {\lower.5ex\hbox{$\sim$}}}\ht1=\grdimen\dp1=0pt
\setbox\simlessbox=\hbox{\raise.5ex\hbox{$<$}\llap
      {\lower.5ex\hbox{$\sim$}}}\ht2=\grdimen\dp2=0pt

\makeatletter 
\renewcommand\@biblabel[1]{}     

\makeatother

\begin{document}

\title[Five WC9 stars discovered in the AAO/UKST \ha\ survey ]
{Five WC9 stars discovered in the AAO/UKST \ha\ survey}
\author[ Hopewell, Barlow, Drew, Unruh, et al  ]
{E. C. Hopewell$^1$, M. J. Barlow$^2$, J. E. Drew$^1$, Y. C. Unruh$^1$, 
Q. A. Parker$^{3,4}$,
\newauthor M. J. Pierce$^5$, P.A.Crowther$^6$, C. Knigge$^7$, S. Phillipps$^5$, 
A. A. Zijlstra$^8$ \\
$^1$Imperial College of Science, Technology and Medicine,
Blackett Laboratory, Exhibition Road, London,  SW7 2AZ, U.K.\\ 
$^2$University College London, Department of Physics \& Astronomy, 
Gower Street, London WC1E 6BT, U.K.\\
$^3$Department of Physics, Macquarie University, NSW 2109, Australia\\
$^4$Anglo-Australian Observatory, PO Box 296, Epping NSW 1710, Australia\\
$^5$Astrophysics Group, Department of Physics, Bristol University, 
Tyndall Avenue, Bristol, BS8 1TL, U.K.\\
$^6$Department of Physics \& Astronomy, University of Sheffield, Hicks Building,
Hounsfield Rd, Sheffield, S3 7RH, U.K.\\
$^7$Department of Physics and Astronomy, University of Southampton, Southampton
 SO17 1BJ, UK.\\ 
$^8$The University of Manchester, School of Physics \& Astronomy, PO Box 88, Manchester M60 1QD, U.K.\\
}


\maketitle
\begin{abstract}
We report the discovery of 5 massive Wolf-Rayet (WR) stars resulting from 
a programme of follow-up spectroscopy of candidate emission line stars in 
the AAO/UKST Southern Galactic Plane \ha\ survey. The 6195-6775 \ang\ spectra 
of the stars are presented and discussed. A WC9 class is assigned to all 5 
stars through comparison of their spectra with those of known late-type WC 
stars, bringing the known total number of Galactic WC9 stars to 44.  Whilst 
three of the five WC9 stars exhibit near infrared (NIR) excesses characteristic
of hot dust emission -- as seen in the great majority of known WC9 stars -- we 
find that two of the stars show no discernible evidence of such excesses. 
This increases the number of known WC9 stars without NIR excesses to 7.    
Reddenings and distances for all 5 stars are estimated. 
\end{abstract}


\section{Introduction}
\label{intro}

Massive Wolf-Rayet stars are important objects of study for
two reasons.  First, the Wolf-Rayet (WR) phase of massive star 
evolution has itself been a significant challenge to models of extreme mass 
loss (Hillier 2003) and represents the likely precursor stage to 
chemically-peculiar core-collapse supernovae (type Ib and Ic: Woosley, Heger
\& Weaver, 2002), which has directed attention toward understanding their 
structure and variety.  Second, their galactic demographics both as a 
function of WR sub-type and, more generally, as markers of star-forming 
activity has also attracted attention in order to achieve a better 
understanding of their stellar evolutionary origins (Maeder \& Meynet 2000) 
and their impact, as luminous objects blowing chemically-enriched, high-speed 
winds into their galactic environments (e.g. Homeier et al 2003).  

WR stars possess very distinctive spectra, dominated by high-contrast, broad 
emission lines.  This reflects the strong mass loss that follows on from the 
loss of the hydrogen-rich atmosphere that these stars were born with, as 
higher mass ($\gtrsim 25$~M$_{\odot}$) O stars. WR stars are most commonly 
H-deficient objects, and so do not normally display prominent hydrogen 
emission lines in their spectra, but are characterised instead by strong 
emission lines of He, N and C.  The equivalent width ratios between the 
various emission lines in WR spectra allow their classification into 
subclasses and subtypes.  WR stars whose spectra are dominated by lines of 
nitrogen and helium are termed WN stars, whilst stars displaying strong 
carbon, oxygen and helium lines are assigned WC or WO classifications. 
Subtypes then refer to the ionisation degree of the star with the WN 3--9 
and WC 4--9 sequences each going from higher to lower ionisation states.

This paper presents the discovery of five Galactic massive Wolf-Rayet stars,
by means of spectroscopic follow-up of the Anglo-Australian 
Observatory United Kingdom Schmidt Telescope (AAO/UKST) \ha\ Survey of the 
Southern Galactic Plane and Magellanic Clouds (Parker et al 2005). These new 
discoveries are located in a relatively small sky area, roughly centred on 
$\ell \sim 339^{\rm o}, b \sim 1^{\rm o}$ and spanning just 4$\times$2 
square degrees. The exceptionally dense young star cluster Westerlund~1 
(Westerlund 1987, Clark \& Negueruela 2002) lies near the edge of this region.
Intriguingly, these 5 objects were the only WR stars discovered in 
spectroscopy of targets within a larger 62 square degree area, which also 
yielded more than 70 new \ha\ emission line stars (Hopewell et al, in prep).
Two of our discoveries lie within the sky area covered by Shara et al (1991, 
1999), in their blue narrowband search for WR stars. 

It is striking that all five new WR stars are shown to belong to the 
same spectral sub-type -- WC9 -- and all are reasonably bright, with 
$R$ magnitudes in the range $14 \lesssim R \lesssim 16$.  The total 
number of known galactic WC9 stars previously stood at 39 (van der 
Hucht 2001: 30, Homeier et al. 2003: 3, LaVine, Eikenberry \& 
Davis 2003: 2, Negueruela \& Clark 2005: 4 in Wd 1)) and as such the 
five discoveries reported here represent a significant addition to 
this sample. The observations revealing them were obtained a 
year after and using the same spectroscopic facility as the discovery 
observations of only the fourth known Galactic WO star, WR 93b 
(Drew et al. 2004).  

The AAO/UKST \ha\ Survey of the Southern Galactic Plane and Magellanic
Clouds was the last photographic UKST sky survey to be carried out. It is
described in full by Parker et al (2005). It was a narrow band photographic 
sky survey of the entire Southern plane of the Milky Way for galactic 
latitudes $-10$\degree $<$ b $<$ + 10 \degree\ and consisted of 233 
Galactic Plane and 40 Magellanic Cloud fields on 4-degree centres. The 
survey was completed in 2003 and is available online, as SuperCOSMOS scans 
of the original survey films at 10 $\mu$m resolution (SHS database, located 
at http://www-wfau.roe.ac.uk/sss/halpha/). This survey used a high 
specification, single element \ha\ interference filter with a 70~\ang\ 
bandpass. This, together with fine-grained Tech-Pan emulsion as the detector, 
allowed a survey of Galactic ionised hydrogen, combining large area coverage 
with good sensitivity and arcsecond resolution.  For each field the aim was 
to take 15-minute broad-band short-red exposures alongside the 3-hour \ha\ 
exposures to produce contemporaneous exposure pairs.  In the SHS, the \ha\
and short-red data are combined with $I$ band data from the older UKST IVN 
Southern Sky Survey.

Since the completion of the survey, a programme of spectroscopic confirmation
of point-source candidate \ha\ emitters from the SHS database has been 
undertaken, using the UKST 6dF multi-object facility. This programme is 
finding all types of \ha\ emitting point sources in the Galaxy, many of which 
represent short-lived evolutionary phases which are, correspondingly, 
rarely observed. Here the focus is on WR stars: their detection via the 
UKST survey and SHS database, typically resulting from the inclusion of the 
strong He~{\sc ii} and/or C~{\sc ii} lines at ${\sim}$~6570~\ang\ within 
the \ha\ filter bandpass. 

The paper is organised as follows.  The next section discusses the method of 
target selection used for the UKST 6dF follow-up spectroscopy program and 
gives magnitudes and positions for the new WR stars. Section 3 then provides 
information about the observations and explains the relevant data reduction 
procedures. The 6dF data, details of spectral subtype identification and 
comments on individual spectra are presented in Section 4. We then consider 
estimates of the dust emission, reddening and distances to the new WR stars 
in Section 5. Finally, we close in Section 6 with a comparison of this red 
selection of WR stars to selections made at other wavelengths, and a brief
comment on how these new stars relate to the previously known WC9 sample.

\section{ Selection of spectroscopic targets from the UKST \ha\ Southern 
Galactic Plane Survey }

\begin{table}
\caption{ 4\degree ${\times}$ 4\degree\ AAO/UKST \ha\ Survey field centers in 
1950 and Galactic co-ordinates.
\label{Field_centers}}
\begin{tabular}{lllll}
\hline
Field   & RA (1950) & Dec (1950) & ${\ell}$ &  b \\
\hline
HAL0348 & 16 00 00 & -48 00 00 & 332.7 & $+$3.3  \\
HAL0349 & 16 24 00 & -48 00 00 & 335.7 & $+$0.5  \\ 
HAL0413 & 16 30 00 & -44 00 00 & 339.3 & $+$2.5  \\  
HAL0414 & 16 52 00 & -44 00 00 & 341.9 & $-$0.4  \\ 
\hline
\end{tabular}
\end{table}

\begin{figure} 
\mbox{\epsfxsize=0.50\textwidth\epsfbox{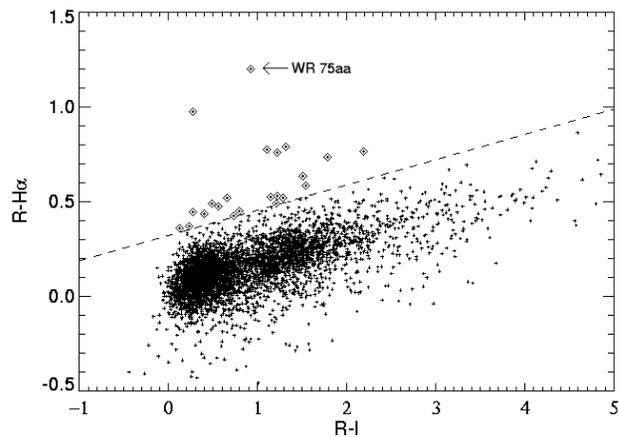}}
\caption{A plot of (R-\ha) versus (R-I) for all SHS catalogue point sources 
within the 1\deg\  box containing WR~75aa which satisfy $14< R < 16.3$ as 
given by the SHS database. The dashed line indicates the selection limit whilst
the diamonds show the sources that were selected for spectroscopic follow-up }
\label{75aa_ccplot}
\end{figure}

The discovery of the five new WC9 stars presented in this paper resulted from
observations of SHS fields HAL0348, HAL0349, HAL0413 and HAL0414, whose field 
centers and positions are shown in Table~\ref{Field_centers}. As would be 
expected for fields straddling the galactic equator, HAL0349 and HAL0414 
are highly reddened regions with \ebv $> 3$ predicted for extragalactic 
objects observed along these sightlines (Schlegel, Finkbeiner \& Davis 1998).
These fields also include a few extremely reddened regions for which 
\ebv $\gtrsim 15$. Fields HA0348 and HAL0413 lie a few degrees higher in 
galactic latitude, and consequently cover regions of less intense reddening 
where the average extinctions are $1 \lesssim $\ebv$ \lesssim 2.5$ and
$1 \lesssim $\ebv$ \lesssim 6$ respectively.

When selecting this region of the galactic plane for follow-up we were
motivated in part by the presence of the star cluster Westerlund 1 in the
corner of HAL0414. Westerlund 1 is a highly reddened young open cluster
whose population has been shown to be rich in massive post-main-sequence
stars (Westerlund 1987; Clark \& Negueruela 2002; Clark et al. 2005).
These studies have revealed a considerable population of supergiants,
hypergiants and Wolf-Rayet stars in the cluster and prompted us to observe
the surrounding region to investigate the environment of this
extraordinary cluster.

\begin{figure*}
\mbox{
\epsfxsize=0.45\textwidth\epsfbox{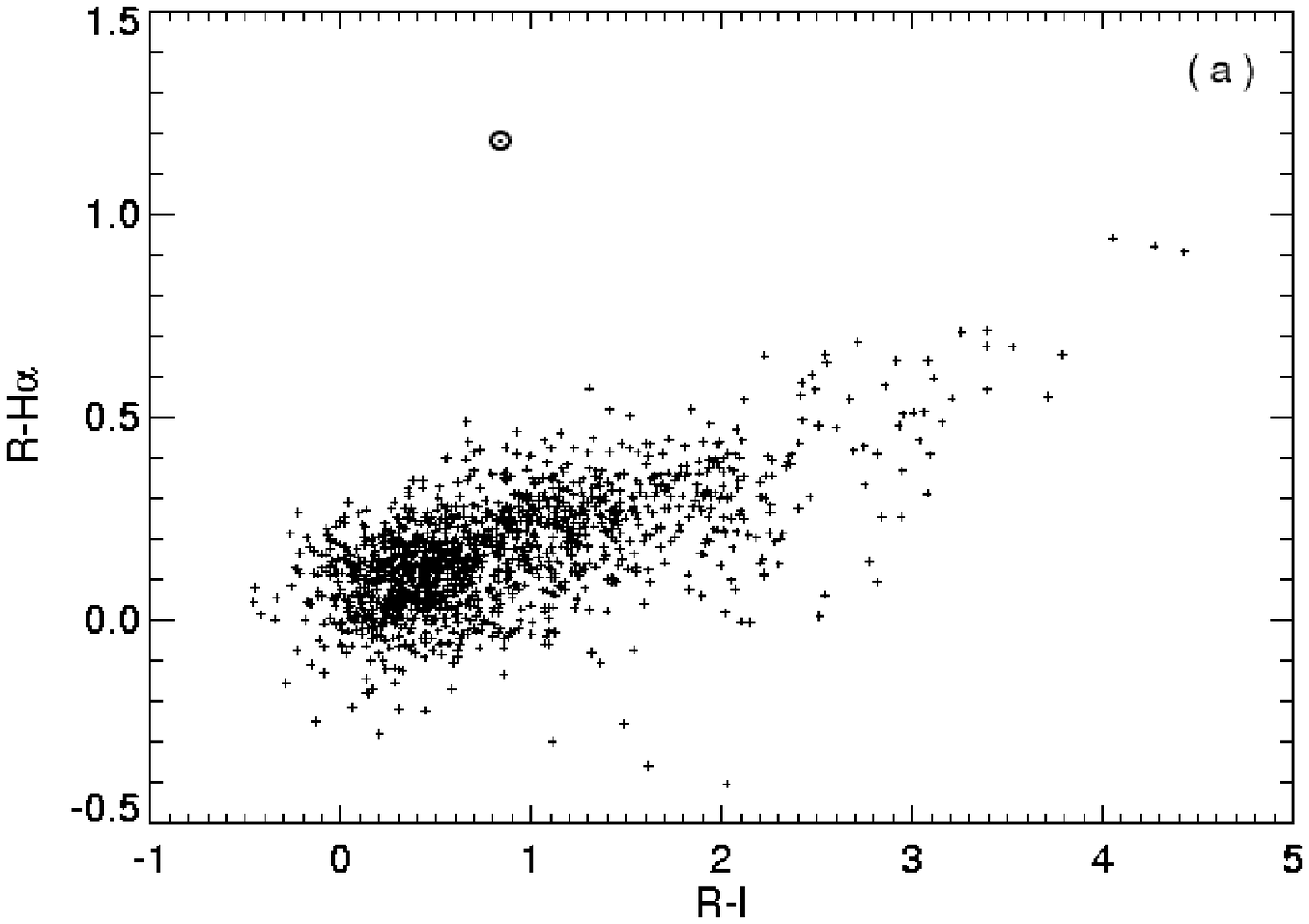}}
\mbox{
\epsfxsize=0.45\textwidth\epsfbox{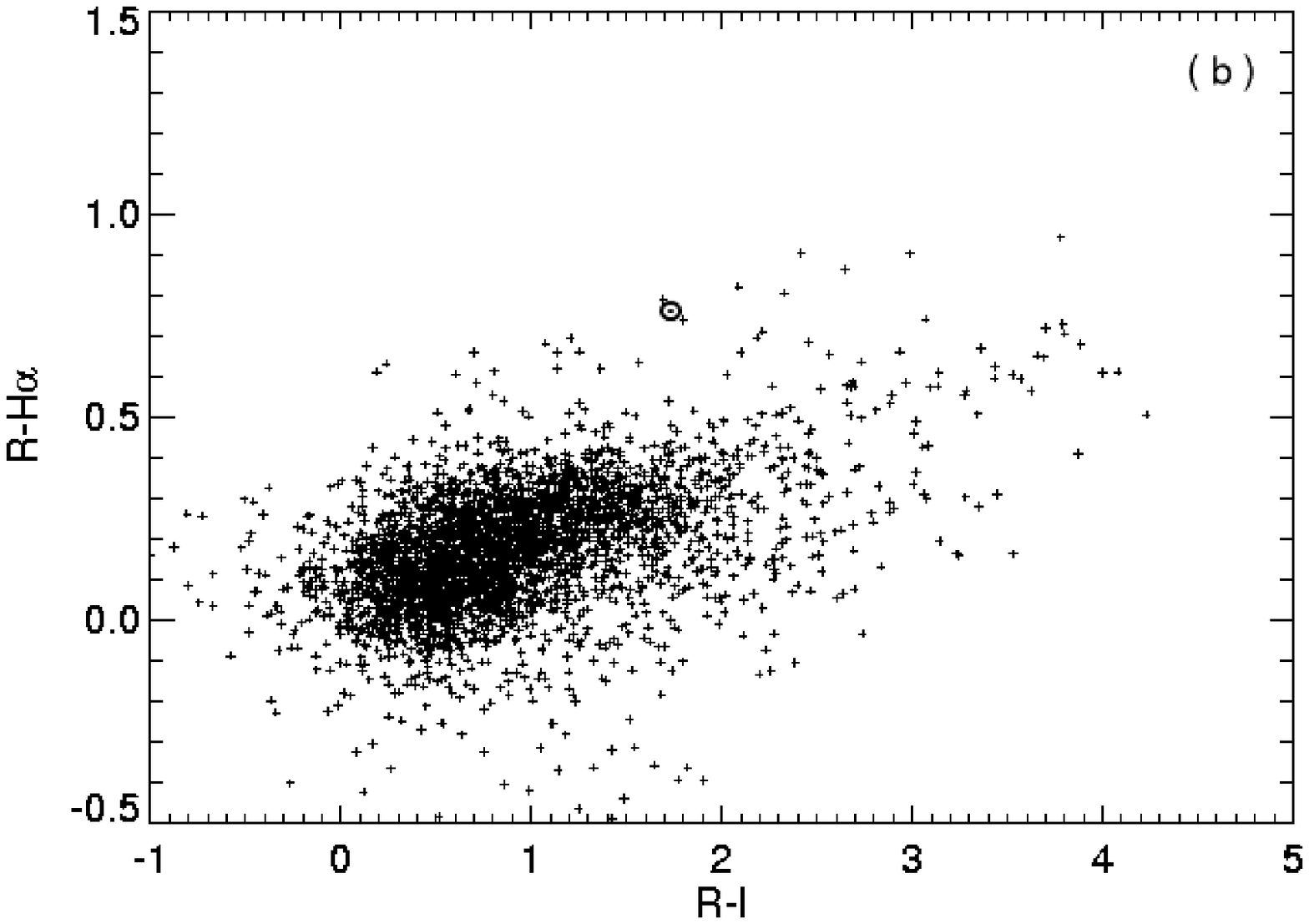}}
\mbox{
\epsfxsize=0.45\textwidth\epsfbox{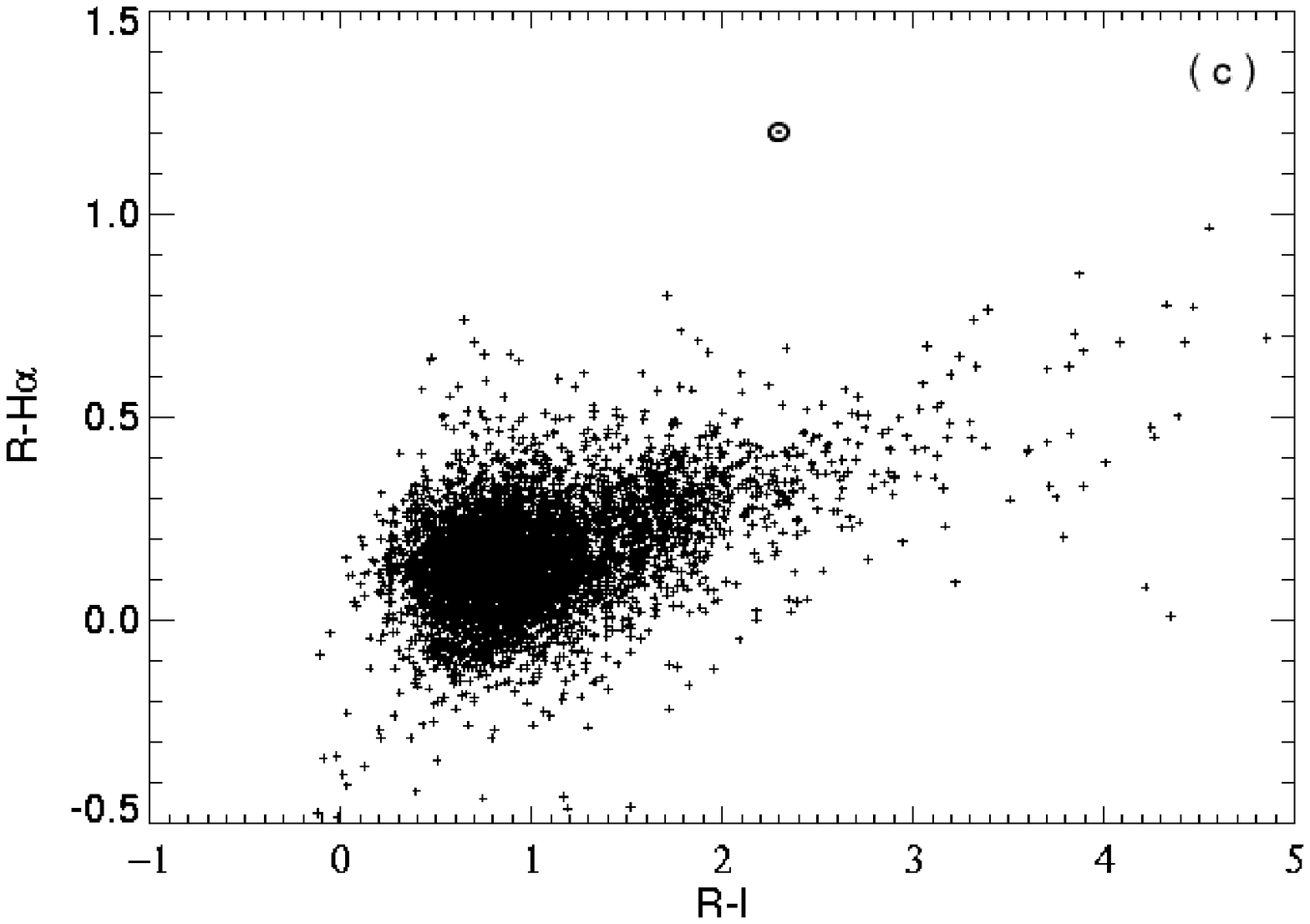}}
\mbox{
\epsfxsize=0.45\textwidth\epsfbox{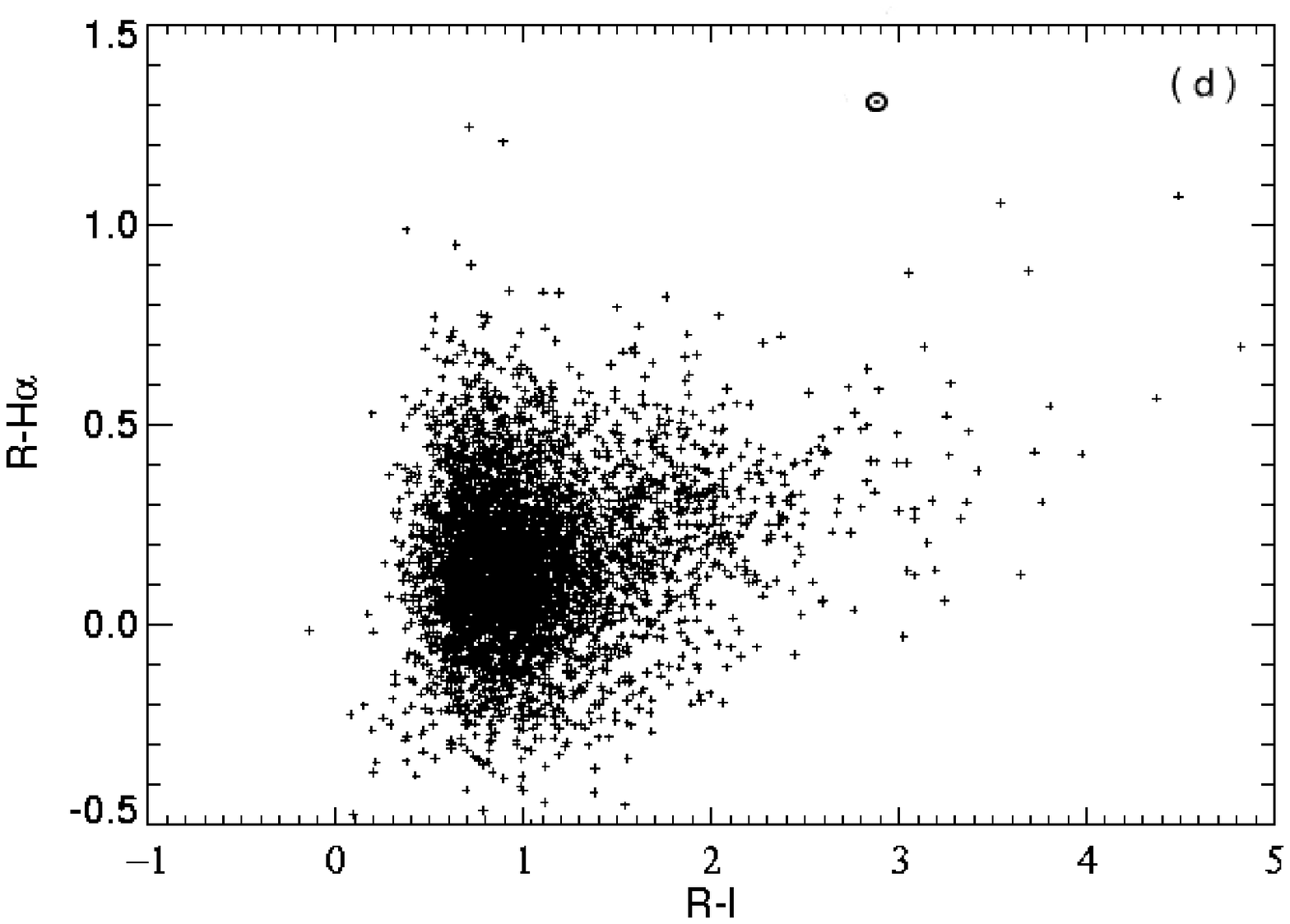}}
\caption{(R-I) versus (R-\ha\ ) for the 1\degree ${\times}$ 1\degree\ ~boxes 
containing (a) WR~75c; (b) WR~75d; (c) WR~77aa; and (d) WR~77t. 
The WR stars are highlighted by open circles.}
\label{cc_plots}
\end{figure*}

\begin{table*}
\caption{Designations (including J2000 positions) and SHS magnitudes for 
the new Wolf-Rayet stars.  The WR catalogue names in the left-hand column
are named according to the conventions of the van der Hucht (2001) catalogue
of massive WR stars.}
\label{WC_init_table}
\begin{tabular}{lllllccc}
\hline
WR No   & SHS designation      & SHS    &  $\ell $   &  $b$  & \multicolumn{3}{c}{SuperCOSMOS Magnitudes}                                              \\
        &                      & Field  &        &         & \ha\    & $R$    & $I$    \\ 
\hline  
75aa    & SHS J162620.2-455946 & 413    & 337.02 & $+$2.17 & 13.85   & 15.31  & 14.18  \\
75c     & SHS J163403.6-434025 & 413    & 339.65 & $+$2.77 & 12.78   & 13.96  & 13.12  \\
75d     & SHS J163417.5-460852 & 349    & 337.86 & $+$1.06 & 13.27   & 14.03  & 12.30  \\
77aa    & SHS J164646.3-454758 & 414    & 339.56 & $-$0.33 & 14.59   & 15.79  & 13.51  \\
77t     & SHS J165057.6-434028 & 414    & 341.66 & $+$0.47 & 14.57   & 15.87  & 13.00  \\
\hline
\end{tabular}
\end{table*}

Target selection was conducted by considering the positions of the objects in a
plot of $(R-I)$ colour versus $(R - H\alpha )$ excess for all SHS point sources 
within a specified $R$ magnitude range. However within these 
$4^{\rm o} \times 4^{\rm o}$ ~fields the location of the stellar locus in this 
colour-colour plot can be seen to change with position in the field, owing to 
variations in the effective magnitude calibration across it. Therefore, when 
identifying the objects for spectroscopic follow-up, the fields were broken 
down into 16 smaller 1\degree$\times$1\degree ~regions and the \ha\ emission 
candidates selected from these subsamples. Two magnitude ranges were considered
separately for selecting targets, $12 < R < 14.5$ and $14.5 < R < 16.3$. 
Figure~\ref{75aa_ccplot} demonstrates how the targets were selected for the area 
that contained WR~75aa, one of the new discoveries. From this sample of 
${\sim}$~5000 stars, 22 were selected as candidates for follow-up spectroscopy,
 all falling between $0 < (R-I) < 3$ and on the \ha\ excess side of the main 
stellar locus. Objects with $(R-H\alpha )$ colours in the excess region, but 
$(R-I) > 3$, were not selected as this region of the plot becomes increasingly 
populated with late type stars, whose optical spectra contain molecular bands, 
which lead to their colours mimicking those of \ha\ excess objects. WR stars 
require ${\sim}$ \ebv\ ${> 5 }$ in order to display such colours, so this 
selection cut also serves to exclude severely reddened stars.

Basic data for the five new Wolf-Rayet stars, including the initial 
SuperCOSMOS magnitudes used to select these objects as targets, are given in 
Table~\ref{WC_init_table}. The R and \ha\ magnitudes in the table refer to the 
AAO/UKST measurements, however the I  magnitudes listed in the SHS catalogue 
were derived through matching SHS sources with $I$ band data from UKST IVN 
Surveys. WR star names have been assigned in line with the convention used 
within the seventh catalogue of Wolf-Rayet stars (van der Hucht, 2001) and 
we will hereafter refer to the new WR stars by these designations.In 
Table~\ref{WC_init_table}, we also introduce the IAU-registered naming 
convention for stars picked out from the SHS: namely, 
SHS~JHHMMSS.s$+$DDMMSS (J2000 coordinates).

From figures~\ref{75aa_ccplot} and \ref{cc_plots} it can be seen that, with
the possible exception of WR~75d, the new WC9 stars were obvious high priority 
candidates for spectroscopic follow-up, displaying the largest $(R-$\ha$ )$ 
excesses in their subfields. WR~75d appears to lie in a more populated region 
of the diagram. However the objects nearby were nearly all discounted for 
follow-up as on closer inspection they were revealed to have spurious \ha\ 
excesses resulting from the confusion of two or more objects in the digitising 
process.

In the case of the candidate which became WR~75aa, the process of checking 
selected targets against known objects revealed that it had already been
identified as a possible \ha\ emission line star. Known as SPH~146, it was 
detected in the southern objective prism survey for \ha\ emission objects 
published by Schwartz, Persson \& Hamann (1977): they described it as showing 
faint/uncertain \ha\ emission.

\section{UKST/6dF Observations }
\label{spec_obs}

\begin{table*}
\caption{Log of SHS 6dF follow-up observations conducted from 26th - 30th March 2004. 
\label{Obs_info}}
\begin{tabular}{|lllllllll|}
\hline
  Field   &  R Mag      & Observation & UT     & Object frame & Sky-offset & Seeing     & Other \\
          &  range      & date        & Start  & exposures    & exposures  &            & information \\
\hline
  HAL0349 & 12-14       & 26-03-2004  & 13:38  & 6 x 1200     & 2 x 1200   & $\sim$ 1$''$ & offset (RA,Dec:-5 -10, +10 +5) \\
  HAL0413 & 12-14       & 28-03-2004  & 13:50  & 3 x 1200     & 1 x 1200   & $\sim$ 2$''$ & offset (RA,Dec:+12 +7)\\
  HAL0413 & 14-16       & 29-03-2004  & 14:50  & 6 x 1200     & 3 x 1200   & $\sim$ 2$''$ & offsets(RA,Dec:+12 -7, -7 +12, +7 +12) \\
  HAL0414 & 14-16       & 27-03-2004  & 15:44  & 6 x 1000     & 2 x 1000   &  2 - 4$''$  & offsets(RA,Dec:-12 -7, -7 +12) \\ 
\hline
\end{tabular}
\end{table*}

Spectroscopic follow-up observations of eight SHS fields were obtained using 
the AAO/UK Schmidt 6dF multi-fibre spectrograph on 26--30 March 2004, three of 
which (HA0349, HA0413 and HA0414) contained the new WR stars presented here.
Details of the observations can be found in Table~\ref{Obs_info}. Observations 
were made using the 1516R grating, which in conjunction with a 
$1024 \times 1024$ EEV detector having 13$\mu$m pixels yields a spectral 
resolution of ${\sim}$ 2~\ang\ and a spectral range of 6195-6775~\ang. 
For the spectra presented in this paper the observations involved taking both 
on-target and sky-offset frames to allow for varying levels of diffuse \ha\ 
emission in the vicinity of our targets.

The data were extracted from the CCD frames and reduced using 6dfdr, the 
6dF adaption of the 2dfdr software package 
(see http://www.aao.gov.au/AAO/2df/manual.htm). Each object and sky-offset 
frame was reduced separately before being combined and then subtracted to 
produce the final spectra. For every frame, this reduction involved: flat field
extraction; fibre-by-fibre arc extraction and calibration; extraction of the 
observed spectrum from each fibre with scattered light correction. Each frame 
was also processed for cosmic ray hits, which the software removed by assuming 
that these corresponded to $>$20$\sigma$ outliers. No fibre throughput 
calibration was applied to these data. Once the individual frames were reduced,
the sky offset frames were reexamined to identify any fibres including unwanted
starlight and, if possible, to replace them with uncontaminated data from a 
different skyframe. No replacement was necessary for any of the offset 
pointings associated with the new Wolf-Rayet stars.  The reduced frames were 
then processed in 6dfdr to produce two combined averaged frames, one for the 
object frames and another for the sky offsets. The sky subtraction was then 
accomplished using FIGARO routines within the Starlink software suite to 
subtract the combined offset frame from the combined object frame, after 
appropriate scaling. This reduction process yielded a single 
wavelength-calibrated frame 
for each field which contained between 80 and 120 spectra, each of 1032 pixels.
As a final measure prior to analysis, the spectra for WR~77aa and WR~77t -- the
faintest of this sample -- were redistributed into 2 pixel bins to reduce 
the noise level.

\section{The 6dF spectra of the 5 WR stars}

\begin{figure} 
\mbox{\epsfxsize=0.50\textwidth\epsfbox{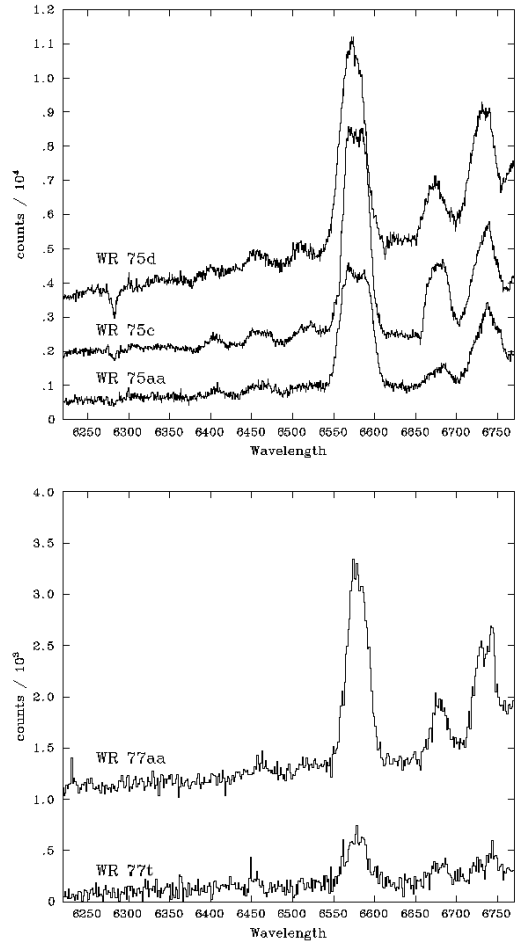}}
\caption{ The 6200~-~6800~\ang\ spectra of the 5 new WR stars, offset 
vertically.   
Upper: WR 75d, 75c and 75aa. Lower: WR 77aa, 77t. 
Note the difference in scale between the two panels.}
\label{WR:spec}
\end{figure}

The extracted stellar spectra are shown in figure~\ref{WR:spec}. It is 
immediately evident that all five stars are of a similar spectral subtype. The 
emission lines present have been identified through use of the Atomic Line List
v 2.04 maintained by P. van Hoof,(http://www.pa.uky.edu/$\sim$peter/atomic/), 
and are listed in Table~\ref{line:list}. All are due to ionised species of 
helium and carbon.

\begin{figure} 
\mbox{\epsfxsize=0.52\textwidth\epsfbox{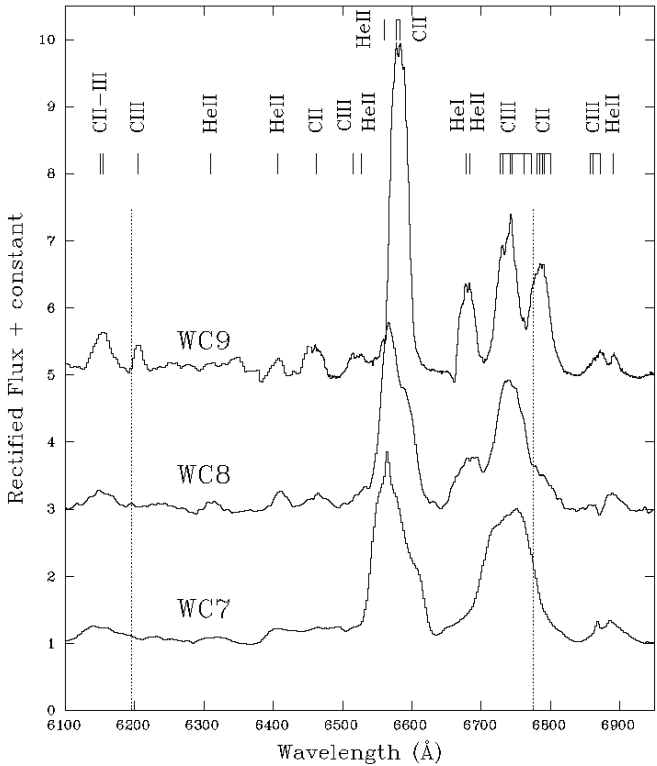}}
\caption{ Example 6200~-~7000~\ang\ spectra of late type WC stars. 
WC7=WR 90, WC8=WR135, WC9=WR103.
The vertical dashed lines indicate the limits of our 6dF spectra.}
\label{wc:late}
\end{figure}

Initial identification of the spectral subtype of these stars was 
accomplished though comparison of the 6dF spectra with the spectrophotometry 
of Wolf-Rayet stars presented by Torres-Dodgen \& Massey (1988). This allowed 
all five to be identified as belonging to the WC9 subtype. 
Figure \ref{wc:late}, illustrates the trends present among late 
type WC stars and confirms our assignments as we now describe.

The spectra shown in figure~\ref{wc:late} illustrate the fact that
as the WR subtype decreases toward earlier types of higher excitation, 
the lines broaden as many features become blends of several lines. 
The strongest emission feature seen in the 6dF spectra, the 6570\ang\ 
blend of He~{\sc ii}~6560~\AA\ and  C~{\sc ii}~6578,82~\AA , is a good 
example of this 
process: The helium line grows in prominence within the blend (towards 
earlier spectral types. Examination of figure~\ref{wc:late} shows that the 
shape of the line profile of the $\sim6570~$\ang\ feature changes markedly 
between subtypes WC8 and WC9. In the WC8 spectrum this feature is very 
obviously a blend, with the C~{\sc ii} 6578,6582~\AA\ doublet weaker 
than the He~{\sc ii} 6561~\AA\ emission, but in the WC9 spectrum the blend 
is dominated by C~{\sc ii}, with He~{\sc ii} 6560~\AA\ virtually absent. 
Since the measured central wavelength of the feature in our 
spectra is 6577.5~\AA\ in four cases and 6573.3~\AA\ in the other (WR 75aa), 
implying C~{\sc ii} dominance, a spectral type of WC9 is implied for all 
five stars.
In support of this assignment, the He~{\sc ii}~6405~\ang\ emission in our 
WR stars is weaker than the adjacent carbon-dominated 6462~\ang\ and 
6515~\ang\ emission features, as seen in the WC9 spectrum shown in 
figure~\ref{wc:late}.

\begin{figure} 
\mbox{\epsfxsize=0.52\textwidth\epsfbox{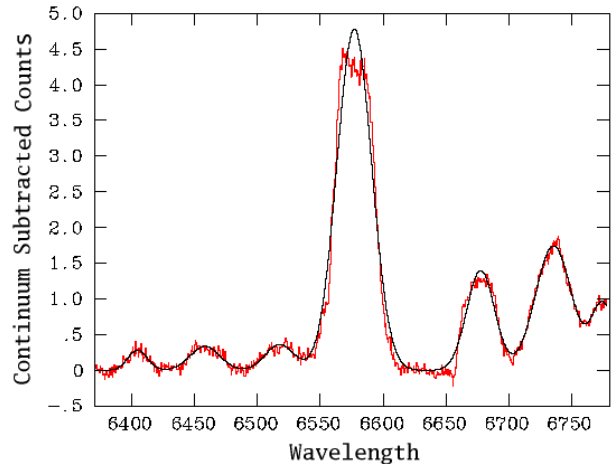}}
\caption{ Example spectral line fitting. The 6375-6775~\ang\ spectrum of WR~75c
(black), with the gaussian fits overlaid (red).}
\label{spec:fit}
\end{figure}

\begin{figure} 
\mbox{\epsfxsize=0.50\textwidth\epsfbox{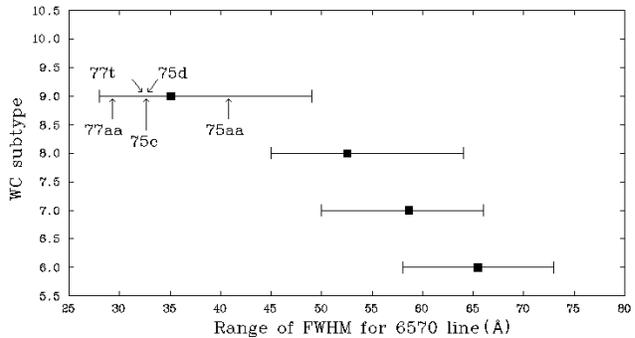}}
\caption{FWHM data for the 6570~\ang\ feature as a function of WC star 
subtype, 
taken from Torres-Dodgen \& Massey's 1988 spectrophotometric atlas. All of 
the WC stars of subtype 6, 7, 8, or 9 in this catalogue were used, provided 
data in the 6500~\ang\ region of their spectra were available. Solid lines 
indicate the FWHM range spanned, filled squares show the mean FWHM for the 
subtype while the FWHMs of the new WC9 stars are indicated with arrows. 
The FWHM were measured via gaussian fitting. }
\label{TD:FWHM}
\end{figure}

The Starlink DIPSO package was used to spectrally fit the lines 
and provide values for equivalent widths (EW) and full widths at half maximum 
(FWHM). For the weaker line EW determinations and all FWHM measurements, 
gaussian fitting was used.  To prepare for this, continuum fits were 
derived for each spectrum and then  divided. This procedure was hampered 
by the narrow spectral range: the red end of the observed spectra lies at 
6775~\ang\ and, as figure~\ref{wc:late} shows, this part of the spectrum is 
crowded with emission lines which make accurate continuum determination 
difficult. For all objects the continuum was estimated separately for the 
regions 6195-6650~\ang\ and 6530-6775~\ang\ using a linear fit in both 
sections.  Figure~\ref{spec:fit} shows an example continuum-subtracted 
spectral line fit for WR~75c: this demonstrates the need to determine the EW 
of the 6570~\ang\ and 6680~\ang\ blends by the more direct device of 
integration of the emission profile counts -- simple gaussian fits 
to these clearly non-gaussian features typically overestimate the line EW.
The EWs for the 6570~\ang\ and 6680~\ang\ blends in Table~\ref{line:list} 
were measured by integrating the net line emission counts after normalising 
the spectrum locally.  It was not possible to fit all the listed emission 
features for all 5 stars: e.g. for the faintest member of the sample, WR~77t, 
gaussians were only fitted to three features.

\begin{table*}
\caption{ Identification and measurements of emission lines in the 
6195-6775~\ang\ spectra of the new WR stars. 
\label{line:list}}
\begin{tabular}{llllllll}
\hline
Wavelength & Transition ID                 & \multicolumn{2}{l}{WR 75aa}       & \multicolumn{2}{l}{WR 75c}        & \multicolumn{2}{l}{WR 75d}      \\
(\ang )    &                               & --EW(\ang )      & FWHM(\ang )    & --EW(\ang )      & FWHM(\ang )    & --EW(\ang )    & FWHM(\ang )    \\
\hline
6405 & He II 15-5                          & ~11.2${\pm}$0.7  & 26.2${\pm}$5.2 & ~~5.7${\pm}$0.9  & 18.7${\pm}$3.5 & ~5.0${\pm}$0.5 & 36.7${\pm}$4.9 \\
6462 & C II 2$s{^2}$6g${\to}$2$s{^2}$4f    & ~20.1${\pm}$1.9  & 36.9${\pm}$4.2 & ~10.4${\pm}$1.1  & 28.5${\pm}$3.4 & ~6.1${\pm}$0.4 & 27.8${\pm}$2.1 \\
6515 & C III 2s9h${\to}$2s6g               & ~12.4${\pm}$1.4  & 26.7${\pm}$3.0 & ~10.8${\pm}$1.1  & 27.3${\pm}$2.6 & ~3.2${\pm}$0.2 & 17.1${\pm}$1.5 \\
     & He II 14-5                          &                  &                &                  &                &                &                \\
6570 & He II 6-4                           & 169${\pm}$10     & 40.8${\pm}$0.4 & 158${\pm}$3      & 32.6${\pm}$0.3 & 74.0${\pm}$0.8 & 32.8${\pm}$0.2 \\
     & C II 3p${\to}$2$s{^2}$              &                  &                &                  &                &                &                \\
6680 & He I 1s3d${\to}$1s2p                & ~17.2${\pm}$5.0  & 29.5${\pm}$2.4 & ~38.7${\pm}$5.0  & 26.7${\pm}$0.4 & 12.8${\pm}$0.9 & 26.6${\pm}$0.6 \\
     & He II 13-5                          &                  &                &                  &                &                &                \\
6735 & C II  M21,M16.03                    & ~75.8${\pm}$2.6  & 35.9${\pm}$1.5 & ~74.8${\pm}$1.1  & 35.2${\pm}$0.6 & 34.7${\pm}$0.4 & 33.5${\pm}$0.4 \\
     & C III M3                            &                  &                &                  &                &                &                \\
\hline
Wavelength & Transition ID                 & \multicolumn{2}{l}{WR 77aa}      & \multicolumn{2}{l}{WR 77t}         &                &                \\
(\ang )    &                               & --EW(\ang )      & FWHM(\ang )    & --EW(\ang )      & FWHM(\ang )    &                &                \\
\hline
6405 & He II 15-5                          &                  &                &                  &                &                &                \\
6462 & C II 2$s{^2}$6g${\to}$2$s{^2}$4f    & ~28.0${\pm}$3.1  & 41.3${\pm}$5.6 & ~10.4${\pm}$2.3  & 9.8${\pm}$2.1  &                &                \\
6515 & C III 2s9h${\to}$2s6g               & ~11.3${\pm}$2.3  & 27.8${\pm}$6.2 &                  &                &                &                \\
    , & He II 14-5                          &                  &                &                  &                &                &                \\
6570 & He II 6-4                           & 178${\pm}$10     & 29.3${\pm}$0.4 & 105${\pm}$10     & 32.2${\pm}$1.6 &                &                \\
     & C II 3p${\to}$2$s{^2}$              &                  &                &                  &                &                &                \\
6680 & He I 1s3d${\to}$1s2p                & 32${\pm}$5       & 19.0${\pm}$2.1 & 32${\pm}$5       & 27.4${\pm}$5.6 &                &                \\
     & He II 13-5                          &                  &                &                  &                &                &                \\
6735 & C II  M21,M16.03                    & ~53.5${\pm}$2.7  & 29.5${\pm}$1.7 &                  &                &                &                \\
     & C III M3                            &                  &                &                  &                &                &                \\
\hline
\end{tabular}
\end{table*}

Some support for the classification of these stars as members of the WC9 class 
comes from applying the same FWHM fitting method to the 6570\ang\ feature in 
the WR spectra in the spectrophotometric catalogue of Torres-Dodgen and 
Massey(1988). Figure~\ref{TD:FWHM} shows that, the later the spectral 
subtype of the star, the smaller the FWHM of the 6570\ang\ emission line blend.
All five of the WC stars in our sample fall within the FWHM range for these WC9
stars, and outside the range shown by the WC8 stars.  WR~75aa is less clear 
cut: it presents with the largest FWHM, as well as the shortest central 
wavelength for this blend -- both suggesting a leaning toward the earlier 
spectral type.

Whilst the five WC9 stars have very similar spectra, there 
are some significant differences between them, which are noted below:

Diffuse interstellar band (DIB) absorption is seen in the spectra of both
WR~75c and WR~75d. For WR~75d the DIBs at 6284~\ang\ and 6613~\ang\ are
both seen, while WR~75c shows evidence of absorption at 6284~\ang\ only. 
The DIB properties will be relevant to the discussion of reddening 
in the next section.

Relatively strong emission in the He~{\sc ii} 6405~\ang\ line is shown by 
WR~75c. The strength of this line compared to the carbon lines is a possible 
indication that this star belongs to the small subgroup of WC9 stars that 
have never shown evidence of circumstellar dust emission. This view derives 
from the work of Williams \& van der Hucht (2000) in which it is shown that 
other WC9 stars of this type (WR 81, 88, 92) also possess relatively 
enhanced He~{\sc ii} emission.

Almost the opposite pattern of EW ratios to that observed for WR~75c is seen 
in the spectrum of WR~77aa. When compared to the other WC9 stars presented 
here, the He{\sc ii} contribution to its spectrum is slight: there is no 
definite detection of He{\sc ii}~6408, while the 6570~\ang\ blend is 
somewhat redshifted and narrower compared to the other WC9 stars presented 
here, suggesting little contribution from He{\sc ii}~6562.

Lastly we note that binarity is suspected for WR~75d on the basis 
that the EW of its 6570~\AA\ emission blend is under half that seen in 
WRs 75aa, 75c and 77aa. This cannot be more than a suspicion since WR 
emission line EWs show significant differences, object to object, even 
among apparently single stars (eg. Torres \& Conti, 1984).

\section{ Dust emission, reddening and distance determination }

\subsection{Reddening estimation}

The next step in assessing the physical properties of the 5 new WR stars 
was to determine the interstellar extinctions towards them and to decide if 
they display NIR colours indicative of circumstellar dust emission. Of the 30
WC9 stars contained within the Seventh catalogue of Galactic WR stars 
(van der Hucht, 2001) only 5 are listed as not possessing circumstellar dust.  
Most of the known WC9 stars (18) are known to display evidence of either
persistent or variable dust emission. The remaining 7 in the catalogue are 
heavily reddened (\av$\sim$29 mag) stars within 30pc of the Galactic Center,
for which it is to difficult reach any conclusion.

\begin{table*}
\caption{ Magnitudes for a selection of known WC9 stars and for all
five newly-discovered WC9 stars.  The $B_J$ and $I$ magnitudes are taken from 
the USNO B1 catalog, the $R$ magnitudes are from the GSC 2.2 catalog, while the
$JHK$ magnitudes are 2MASS measurements.We also quote approximate MSX A-band 
(8.28~$\mu$m) magnitudes, $m_{8.28}$, where available. Exceptions are the 
R mags for WR 103 - taken from the USNO catalog and the I mags of WR 119 
and 77t taken from the DENIS catalogue.
\label{extmag}}
\begin{tabular}{lrrrrrrr}
  &         &                        &                        &                        &                        &                        &              \\
\hline
 WR  &\multicolumn{1}{c}{$B_J$} &\multicolumn{1}{c}{$R$} &\multicolumn{1}{c}{$I$} &\multicolumn{1}{c}{$J$} &\multicolumn{1}{c}{$H$} &\multicolumn{1}{c}{$K$} &  $m_{8.28}$  \\
     &                          &                                                 &                        &                        &                        &              \\
\hline 
\multicolumn{5}{l}{Known WC9 Stars without NIR dust excesses}                                           &                        &                        &              \\
75a  & 16.2                     & 14.1${\pm}$0.4         & 11.96${\pm}$0.3        & ~9.96${\pm}$0.02       & ~9.18${\pm}$0.03       & ~8.50${\pm}$0.02       &              \\
75b  & 15.8                     & 14.1${\pm}$0.4         & 11.65${\pm}$0.3        & ~9.76${\pm}$0.03       & ~9.00${\pm}$0.03       & ~8.36${\pm}$0.03       &              \\
81   & 13.2                     & 11.0${\pm}$0.5         & 10.18${\pm}$0.3        & ~8.29${\pm}$0.02       & ~7.76${\pm}$0.05       & ~7.12${\pm}$0.02       &   5.8        \\
88   & 14.1                     & 11.6${\pm}$0.4         & 10.62${\pm}$0.3        & ~9.03${\pm}$0.02       & ~8.56${\pm}$0.04       & ~8.05${\pm}$0.04       &   6.7        \\
92   & 10.5                     & 10.2${\pm}$0.1         & 10.12${\pm}$0.3        & ~9.50${\pm}$0.03       & ~9.22${\pm}$0.03       & ~8.82${\pm}$0.02       &              \\
\hline 
\multicolumn{5}{l}{Known WC9 Stars with NIR dust excesses}                     &                                          &                        &              \\
65   & 14.4                     & 12.3${\pm}$0.4         & 10.96${\pm}$0.3        & ~8.46${\pm}$0.02       & ~7.28${\pm}$0.05       & ~6.17${\pm}$0.03       &   4.5        \\
73   & 15.5                     & 13.5${\pm}$0.3         & 11.81${\pm}$0.3        & 10.32${\pm}$0.02       & ~8.79${\pm}$0.05       & ~7.47${\pm}$0.02       &   4.8        \\
95   & 14.5                     & 12.3${\pm}$1.1         & 10.92${\pm}$0.3        & ~8.29${\pm}$0.02       & ~6.67${\pm}$0.03       & ~5.27${\pm}$0.02       &   2.4        \\
103  & ~8.9                     & ~8.7${\pm}$0.3         & ~8.64${\pm}$0.3        & ~7.75${\pm}$0.03       & ~7.21${\pm}$0.05       & ~6.37${\pm}$0.03       &   4.3        \\
104  & 13.9                     & 12.1${\pm}$0.4         & 10.31${\pm}$0.3        & ~6.67${\pm}$0.03       & ~4.34${\pm}$0.24       & ~2.42${\pm}$0.26       &  -1.9        \\
119  & 13.3                     & 11.3${\pm}$0.4         & 10.99${\pm}$0.3        & ~9.50${\pm}$0.02       & ~8.43${\pm}$0.06       & ~7.27${\pm}$0.02       &   4.8        \\
\hline  
\multicolumn{4}{l}{New WC9 Stars}                     &                        &                        &                        &              \\
75aa & 17.5                     & 15.5${\pm}$0.2         & 14.35${\pm}$0.3        & 12.03${\pm}$0.02       & 10.73${\pm}$0.02       & ~9.46${\pm}$0.02       &   6.9        \\
75c  & 16.4                     & 14.2${\pm}$0.2         & 13.11${\pm}$0.3        & 11.63${\pm}$0.02       & 11.12${\pm}$0.02       & 10.52${\pm}$0.02       &              \\
75d  & 17.0                     & 14.8${\pm}$0.2         & 11.99${\pm}$0.3        & 10.68${\pm}$0.02       & ~9.88${\pm}$0.02       & ~9.12${\pm}$0.02       &              \\
77aa & 19.3                     & 16.0${\pm}$0.2         & 13.54${\pm}$0.3        & 10.04${\pm}$0.02       & ~8.21${\pm}$0.02       & ~6.73${\pm}$0.03       &   4.0        \\
77t  & 18.9                     & 16.1${\pm}$0.5         & 13.85${\pm}$0.4        & 10.65${\pm}$0.03       & ~9.41${\pm}$0.02       & ~8.32${\pm}$0.02       &   6.0        \\
\hline
     &      &                &                &             &                   &                    &                 \\
\end{tabular}
\end{table*}

To estimate the reddening of our WC9 stars, data from the 2MASS and GSC 2.2
catalogues were obtained for 11 well studied WC9 stars and the five newly 
discovered objects (Table~\ref{extmag}). We split the 11 known WC9 stars 
into the group of five without NIR excesses and a group of six with 
excesses. We chose to use the 2MASS $JHK$ and GSC 2.2 $R$ 
magnitudes to derive measures of the spectral energy distributions (SEDs) of 
our sample because reasonably uniform data on all 16 WR stars are contained 
within these catalogues. Table~\ref{extmag} gives the relevant magnitudes.

The initial assumption of our method for determining reddenings is that all 
WC9 stars have similar unreddened optical/NIR colours, allowing the better
known examples to serve as references SEDs for the newly discovered objects. 
More specifically, we dereddened the magnitudes of the five WC9 stars without 
NIR dust excesses, in order to obtain their intrinsic colours which were 
averaged to use as template colours.  In the process we noted that WR 92 
is somewhat discrepant, but not so much that its inclusion significantly 
altered the derived SED template. Initially, we considered just red
and infrared wavelengths.  For these stars we adopt the \av\ 
values from Table 28 of the seventh catalogue of WR stars (van der Hucht, 2001)
and used the tabulation of \alam/\av\ of Cardelli, Clayton \& Mathis (1989)
to derive dereddened magnitudes with a standard Galactic reddening law, 
corresponding to R=3.1, applied. 

To estimate the reddenings towards other stars in our sample, we dereddened 
their observed NIR magnitudes until the best match to our adopted dereddened 
NIR colours was achieved. In this process, we attached the highest weight to 
matching $(J-K)$. However, this method can only give reliable reddening 
estimates for the minority of WC9 stars without dust emission. When instead, 
there is significant contamination of the NIR SED by dust emission, this 
approach results in an anomalously high derived visual extinction. The listed 
$A_{V,(K)}$ errors in Table~\ref{reddust} take no account of this systematic
difficulty. They indicate only the results of propagating through the errors 
inherent to the photometry.

Overestimation of $A_{V,(K)}$, due to the contamination of, especially, 
$K$-band light by warm dust emission, can be put to use as a useful 
pointer to NIR dust emission. To do this  and to test 
the validity of the $A_{V,(K)}$ values deduced from matching the NIR colours, 
we used our derived intrinsic WC9 $(R-K)$ colour, first to obtain dereddened 
$R$ magnitudes from the dereddened $K$ magnitudes for all 16 WC9 stars in the
sample, and second, `predicted' reddened $R$ magnitudes (reddened by amounts 
consistent with the 2MASS NIR colour fitting).  In the absence of NIR dust 
emission these predicted magnitudes, $R_{\rm (K)}$, should roughly match 
observed $R$ magnitudes taken from the GSC 2.2 catalogue.  For stars with NIR 
dust emission, this procedure yields clearly discrepant, fainter magnitudes 
than those observed.  Hence, objects with large $(R - R_{\rm (K)})$ are picked 
out as NIR excess objects.  The results of applying this check to our sample 
are given in Table~\ref{reddust}. In this way we successfully reproduce the 
division of the known WR stars into the NIR-excess and no-excess categories, 
with the exception of WR 92 and WR 103 which cannot be assigned to either 
category solely on the basis of their $(R - R_{\rm (K)})$ values. 
These are 'intermediate' cases where  $(R - R_{\rm (K)})\sim 1$.
For the newly discovered WC9 stars, we find that both WR~75c and WR~75d appear 
to be without NIR dust emission, increasing the number of recognised Galactic 
`non-dusty' WC9 stars from 5 to 7. In WR 77t, $(R - R_{\rm (K)})$ is 
${\sim}$~1 making it intermediate such that we cannot exclude or confirm 
the presence of a NIR dust excess.

\begin{table*}
\caption{Reddening and distance estimates. The column headed `dust?' notes 
whether NIR excesses attributable to dust emission are present.  The following 
columns give derived reddenings and related quantities: $A_{V,(cat)}$ is the 
visual extinction given in Table 28 of the seventh catalogue of galactic 
WR stars (van der Hucht, 2001); $A_{V,(K)}$ is the extinction we derive using 
the 2MASS $JHK$ data; $R-R_{(K)}$ is the difference between the GSC 2.2 $R$ 
magnitude and the predicted R magnitude based on the NIR-based reddening 
estimate; $A_{V,(B)}$ is the extinction obtained from the $(B_J-J)$ colour; 
$D_{cat}$ is the distance also given in Table 28 of the WR catalogue (van der 
Hucht (2001); $D$ is the estimated distance, and was used in conjunction with 
the Galactic co-ordinates to estimate $R_G$, the galactocentric radius.  The 
data in these columns were derived using a reference SED determined from the 
mean dereddened colours of the 5 non-dusty WC9 stars WR 75a, 75b, 81, 
88 and 92, adopting $A_V$ values from van der Hucht(2001).
\label{reddust}}
\begin{tabular}{lrrrrrrrr}
     &       &                  &                &             &                    &           &                 &                 \\
\hline
WR   & dust? & $A_{V,cat}$      & $A_{V,(K)}$    & $R-R_{(K)}$ &  $A_{V,(B)}$       & $D_{cat}$ & $D$             & $R_G$           \\
     &       &                  &                &             &                    & (kpc)     & (kpc)           & (kpc)           \\
\hline  
\multicolumn{6}{l}{Known WC9 stars without NIR dust excesses}                       &           &                 &                 \\
75a  & no    &  8.1            &  8.6${\pm}$0.6  & -0.1        &   8.2${\pm}$0.4    &    1.6    & 3.3${\pm}$0.5   &                 \\
75b  & no    &  8.9            &  8.2${\pm}$0.7  & -0.5        &   7.9${\pm}$0.4    &    2.3    & 3.6${\pm}$0.6   &                 \\
81   & no    &  6.4            &  6.9${\pm}$0.5  &  0.5        &   6.6${\pm}$0.4    &    1.6    & 1.4${\pm}$0.2   &                 \\
88   & no    &  6.0            &  5.8${\pm}$0.7  &  0.1        &   6.7${\pm}$0.4    &    2.3    & 1.7${\pm}$0.3   &                 \\
92   & no    &  2.1            &  4.0${\pm}$0.6  &  1.2        &   1.8${\pm}$0.4    &    3.8    & 5.0${\pm}$0.8   &                 \\
\hline 
\multicolumn{6}{l}{Known WC9 stars with NIR dust excesses}                          &           &                 &                 \\
65   & yes   &  7.6            & 13.4${\pm}$0.6  &  2.4        &   7.8${\pm}$0.4    &    3.3    & 1.6${\pm}$0.3   &                 \\
73   & yes   &  6.9            & 16.6${\pm}$0.5  &  4.6        &   6.9${\pm}$0.4    &    3.9    & 3.8${\pm}$0.6   &                 \\
95   & yes   &  7.4            & 17.6${\pm}$0.4  &  4.2        &   8.2${\pm}$0.4    &    2.1    & 1.4${\pm}$0.2   &                 \\
103  & yes   &  1.8            &  8.1${\pm}$0.7  &  1.6        &   2.0${\pm}$0.4    &    2.2    & 2.3${\pm}$0.4   &                 \\
104  & yes   &  7.2            & 24.7${\pm}$1.7  &  6.1        &   8.7${\pm}$0.4    &    2.3    & 1.1${\pm}$0.2   &                 \\
119  & yes   &  4.4            & 13.1${\pm}$0.5  &  4.3        &   5.3${\pm}$0.4    &    3.3    & 2.5${\pm}$0.4   &                 \\
\hline 
\multicolumn{6}{l}{New WC9 stars}                                                   &           &                 &                 \\
75aa & yes  &                   & 15.0${\pm}$0.5 &  3.6        &  7.3 ${\pm}$0.4    &           &  8.4${\pm}$1.3  & 3.4${\pm}$0.8   \\
75c  & no   &                   &  6.5${\pm}$0.5 &  0.5        &  6.4 ${\pm}$0.6    &           &  6.2${\pm}$1.4  & 3.5${\pm}$1.7   \\
75d  & no   &                   &  9.1${\pm}$0.5 &  0.2        &  8.3 ${\pm}$0.4    &           &  4.3${\pm}$0.7  & 4.8${\pm}$1.4   \\
77aa & yes  &                   & 19.3${\pm}$0.6 &  3.1        & 11.9 ${\pm}$0.4    &           &  2.2${\pm}$0.4  & 6.5${\pm}$1.0   \\
77t  &      &                   & 13.6${\pm}$0.6 &  0.9        & 10.3 ${\pm}$0.4    &           &  3.9${\pm}$0.6  & 4.9${\pm}$1.5   \\
\hline
\end{tabular}
\end{table*}

The problem, for extinction estimation, of dust emission contaminating
the NIR SEDs of WC9 stars can be overcome by using shorter wavelength
magnitudes that are uncontaminated. To this end, the $B_J$ magnitudes of the 
stars in our sample given in the USNO B1.0 catalogue were collected, and are 
believed to be accurate to ${\pm}$ 0.3 magnitudes (Monet et al. 2003). 
Essentially the same reddening determination method was then applied to all 
stars in our sample, with the difference that all stars were dereddened to a 
common intrinsic $(B_J - J)$ value, in place of $(J-K)$. The $(B_J - J)$ 
colour index was chosen for this purpose  because it covers a region of the
SED which we would expect to be far less affected by dust emission than (J-K)
whilst spanning a wide wavelength range. Since the photographic $B_J$ bandpass
is a broad one and these WC9 stars are significantly reddened, we re-used
the group of five known WC9 stars without NIR excesses, this time to serve as 
templates to establish a best estimate for the effective wavelength of the
$B_J$ bandpass.  We found this to be $\sim$4800~\AA , fixing \abj/\alam\ to 
be 1.1: these give a mean intrinsic $(B_J - J)$ colour of -0.5 for the
catalogue reddenings of these five objects.  Resulting estimates of 
extinctions ($A_{V,(B)}$) calculated on this basis for all the WC9 stars are 
listed in Table~\ref{reddust}. The errors on these estimates are again from 
propagation of photometric error only. 

For the already known WR stars there is now reasonable, if less precise, 
agreement between our new estimated extinctions, $A_{V,(B)}$, and the 
already published catalogue values ($A_{V,(cat)}$): the mean difference is
$\sim$0.5 magnitudes.  As a necessary result of the method adopted, the
agreement in the mean is best for the known WC9 stars with no apparent dust 
excess, since these provided the mean template $(B_J - J)$ colours.  There
is evidence of a positive offset for the known WC9 stars showing NIR dust 
emission, in the sense that $A_{V,(B)}$ always exceeds $A_{V,cat}$.  This 
excess is most noticeable for WR~104.  We attribute this to dust emission
being a greater contaminant of the $J$ band flux in this object.  
Nevertheless it is clear that $A_{V,(B)}$ is a very much better measure of 
the visual extinction for objects with dust emission than $A_{V,(K)}$.

We conclude that for WR~75c and 75d, without NIR continuum excesses, that
$A_{V,(K)} = 6.1\pm0.5$ and $8.8\pm0.5$ are to be preferred as extinction
estimates.  For WR~75aa, 77aa and 77t, in which there is evidence of dust
emission, the more approximate estimates exploiting blue photographic 
magnitudes are to be preferred: namely, $A_{V,(B)} = 7.3\pm0.7$, $11.9\pm0.7$ 
and $10.3\pm0.7$.

\subsection{Distance estimates, observed SEDs and other data}

\begin{figure*} 
\mbox{\epsfxsize=0.75\textwidth\epsfbox{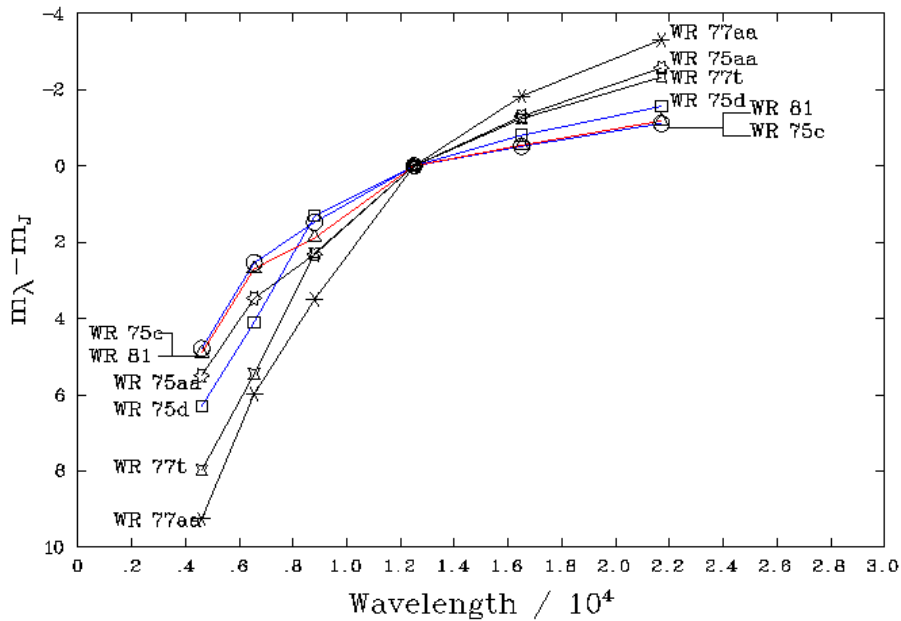}}
\caption{The observed SEDs of the newly discovered WC9 stars.  For comparison,
 the data on the non-dusty star WR 81 are also plotted.  The ordinate 
is the broadband magnitude difference, $m_{\lambda} - m_J$, and is derived 
from the data given in table~\ref{extmag}, and the $I$ magnitudes given in the 
USNO B1 catalogue. At NIR wavelengths, WR 75c and WR 75d (shown in blue) are 
most like WR 81 (shown in red) even before dereddening.}
\label{seds}
\end{figure*}

The derived extinctions, $A_{V,(B)}$, can be combined with the tabulated 
GSC~2.2 $R$ magnitudes and the assumption ${M_V}={M_R}=-4.6$ (using the value 
of $M_V$ from van der Hucht 2001), in order to estimate distances to the 
newly-discovered WC9 stars.  These are given in the penultimate column of 
Table~\ref{reddust}.  The errors specified are likely to be optimistic
given that the distances calculated for the known WC9 stars differ
from their catalogued values by 1--2~kpc -- differences larger than 
the errors deriving from the photometric uncertainties alone. The 
probable causes of this scatter are the limited reliability of the individual
GSC 2.2 $R$ magnitudes and uncertainties in \av\ due to possible differences 
between the assumed and actual reddening law for different objects.
In addition, among the known WC9 stars with NIR
excesses, there is evidence of a systematic effect such that our estimated 
distances are always too low.  This can be explained by the $A_{V,(B)}$ 
estimates being typically higher than they are in reality (as we suspect, see 
above in Section 5.1).  We conclude that the distance estimates to the
five newly-discovered WC9 stars are rough values only -- more reliable CCD
photometry can certainly improve the situation.  For WRs 75aa, 77aa and
77t there are further grounds to suspect our distance estimates are on the
low side.

In figure~\ref{seds} we plot the optical and NIR colours we have gathered 
from the literature on the newly discovered objects. It confirms that 
WR 75c and WR 75d are, in the NIR at least, the most similar in colour 
to the non-dusty WC9 stars.  Because of this and the success with which we 
are able to `predict' the GSC 2.2 $R$ magnitude, we are confident we have 
measured the visual extinctions of WR 75c and WR 75d in a self-consistent
way.

We have checked for MSX (Egan \& Price 1996) detections of mid infrared
(8.28~$\mu$m) emission at or near the positions of our sample WC9 stars.
This shows the expected correlation between detectable NIR excesses and
mid-IR flux: the 3 new WC9 stars deduced to have NIR excesses were
detected by MSX, while the two without excesses (WR 75c and 75d) were not.
The $m_{8.28}$ magnitudes included in Table~\ref{extmag} are based on a
zero magnitude flux from Vega at 8.28~$\mu$m of 55~Jy, derived by 
normalising a Kurucz T$_{\rm eff}$~=~9400~K, log~g~=~3.90 model to Vega's
5556~\AA\ flux of 3540~Jy. The IR emission of the newly-discovered WC9 
stars is fairly faint, which should not be a surprise given that they escaped 
detection by previous IR colour-based searches for dusty Wolf-Rayet stars 
(e.g. Cohen's 1995 IRAS search).

As mentioned in section 4, both of the new WC9 stars lacking NIR dust excesses 
display DIB absorption features in their spectra. The EWs of these DIBs can be 
compared to those of HD~183143, the standard DIB reference object (Herbig 
1995), in order to obtain a rough lower limit to the interstellar reddening. 
The EW and FWHM values of the DIB features were measured using the same 
spectral fitting procedure as was applied to the emission lines, and the 
results are given in Table~\ref{dibs}. The $A_{V,(DIB)}$ values for WR~75c 
and 75d resulting from such a comparison are considerably 
lower than those estimated from their NIR colours in Table~\ref{reddust}.  
This suggests that about half of the reddening towards these stars may be 
associated with molecular gas -- not commonly thought to host the DIB 
carriers -- rather than with diffuse-cloud atomic gas that does.
(see Herbig 1995 and references therein). 

\begin{table}
\caption{ Identification and measurement of the observed DIB features. 
${\lambda_{ref}}$ and $_{ref}$ refer to the properties of HD 183143.
\label{dibs}}
\begin{tabular}{cccrrc}
\hline
${\lambda_{ref}}$ & ${\lambda_{obs}}$ & FWHM & EW$_{ref}$ & EW$_{obj}$ 
   & $A_{V,(DIB)}$  \\
 (\AA ) & (\AA ) & (\AA ) & (m\AA ) & (m\AA ) & \\
\hline 
WR 75c:            &   &  &  &  \\
6283.86 & 6282.0 & 6.2 & 1945 & 1600 & 3.3  \\
\hline
WR 75d:  &          &   &  &  &  \\
6283.86 & 6282.2 & 8.5 & 1945 & 3400 & 6.9  \\
6613.62 & 6612.5 & 2.8 & 358  & 500  & 5.5  \\
\hline
\end{tabular}
\end{table}

The coordinates of WR 77aa place it within 5~arcmin of Westerlund 1, the
star cluster which first motivated our investigation of this region.
WR~77aa has the largest reddening (A$_{\rm V} = 11.9\pm0.7$ mags) of the
five WC9 stars discussed here, not far short of the A$_{\rm V}$ = 13.6
estimated by Clark et al. (2005) for a number of OB supergiants at the
core of Wd~1, for which Clark et al. estimate a distance of 2~--~5.5~kpc. The
distance of 2.2$\pm$1.4 kpc derived here for WR 77aa opens the possibility
of association with Wd~1. 

\section{Discussion}

The allocation of a WC9 class to all five of the WR stars discussed here is 
consistent with the already known galactic distribution of this class of 
object. The 5 WC9 stars presented here fall within 
$337.0^{\rm o} < \ell < 342^{\rm o}$ and $-0.35^{\rm o} < b < 2.80^{\rm o}$ 
with galactocentric radii in the range $3.4 \lesssim R_{G} \lesssim 6.5$~kpc
(Table~\ref{reddust}). This finding is based on distance estimates that are 
presently uncertain by as much as $\sim$2~kpc.  Once CCD photometry becomes 
available for these newly-discovered WC9 stars, these errors will fall.  

\begin{figure}
\mbox{\epsfxsize=0.52\textwidth\epsfbox{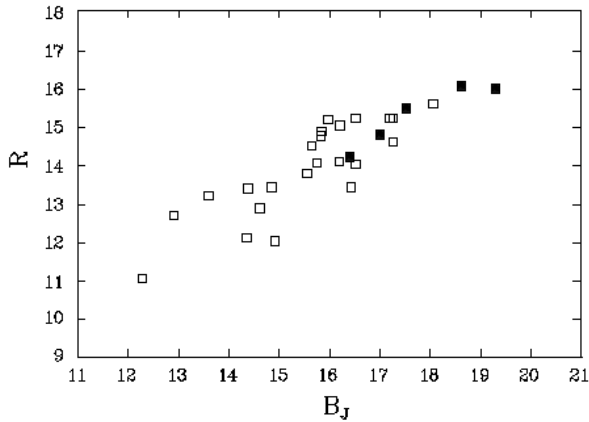}}
\caption{ A plot of GSC 2.2 $R$ magnitudes against USNO $B_J$ magnitudes 
for the WR stars detected by the Shara et al (1999) survey of the southern 
Galactic Plane (open squares) and the newly discovered WC9 stars presented in 
this paper (filled squares). Only 27 of the 35 WR stars detected by Shara 
et al. are plotted, as not all stars in their sample had reliable data 
available from both the GSC 2.2 and USNO catalogues.}
\label{sharamags}
\end{figure}

The small region encompassing the five new Wolf-Rayet stars partially overlaps 
the area covered by the Shara et al. (1991, 1999) survey for galactic WR 
stars, with WR 75aa, 75d and 77aa all falling within the overlap region. 
Their survey used the comparison of narrow and broadband photometry centred 
on the 4686~\ang\ WR emission line feature to select WR candidates for 
spectroscopic observation. It probed down to a blue magnitude of ${\sim}19$ 
and was considered to be complete for the detection of single WR stars at the 
90\% level down to ${\sim}$17.5.  The broadband magnitudes listed by Shara et 
al. for the stars discovered by their survey tend to be fainter than the 
$B_J$ magnitudes for the same stars given in either the whole-sky GSC 2.2 or 
USNO B1.0 catalogues. This no doubt reflects the difference between effective 
$B$ mean wavelengths, which are somewhat shorter in the Shara et al 
measurements. However, as USNO B1.0 magnitudes are also available for the new 
discoveries we need to use them in comparing the $B_J$ and $R$ magnitudes of 
the new WC9 stars with those of the Shara et al sample (Fig~\ref{sharamags}).  
This comparison demonstrates that the new Wolf-Rayet stars selected from the 
SHS, with $R$ magnitudes in the range $14 < R < 16$, overlap with and reach 
beyond the faint end of the Shara et al survey.

The Shara et al (1991) survey was the last large scale optical survey for 
galactic WR stars. All other galactic WR star discoveries since 1991 have 
either been the result of the reclassification of previously known emission 
line objects or resulted from NIR/MIR/radio observations. Longer wavelengths 
suffer far less severely from the effects of extinction, and so allow the 
detection of WR stars in distant parts of the galaxy obscured by large columns 
of dust.  Recent WR star searches have therefore focused on $K$ band (2\mic ) 
observations. In such searches there is a potential bias toward the discovery
of WC9 stars with NIR continuum excesses as they are brighter at NIR
wavelengths.  But, at the same time emission line EWs in WC9 stars NIR spectra 
may be reduced by the added dust continuum, and therefore become harder 
to detect.

In order to estimate the frequency of occurrence of dust emission in late WC 
stars, a sample that is flux-limited at wavelengths shortward of 1~$\mu$m  
could help avoid the biases that might be acting at longer wavelengths. Of the 
five newly-discovered WC9 stars presented here, selected via their strong 
C~{\sc ii} $\lambda\lambda$6578,6582 emission, two have turned out not to 
possess NIR excesses attributable to dust.  Up to now, just five -- out of 
the more than 30 known WC9 stars --  have been recognised as belonging to this 
category.  Three of them were discussed by Williams \& van der Hucht (2000). 
The other two, WR~75a and 75b, were products of the Shara et al survey, and
were recognised as lacking NIR excesses by van der Hucht (2001). So all seven 
objects, making up this group, are optical discoveries.

It was noted a long time ago (e.g. Smith 1968) that Galactic WC9 stars are
found only inside the Solar Circle.  Despite the large increase in the 
numbers of known Galactic WC9 stars since then, this conclusion still holds.
In the lower metallicity Large Magellanic and Small Magellanic Clouds, no
WC9 stars at all are known\footnote[1] BAT99-4 is listed as type WC9+O8V 
in the Breysacher et al (1999) catalogue of WR stars in the LMC, but it is 
noted that this typing had been dismissed by Moffat (1991). Heydari-Malayeri
\& Melnick (1992) confirmed that the older WC9 assignment was not warranted. .
In contrast to this, M83 -- a metal rich spiral 
like the Milky Way -- has been shown to host a large number of late WC stars, 
including some WC9s (Hadfield et al, 2005). 
Another aspect to this is the recognition that there is a gradient in 
the number of WC type stars relative to WN stars such that the WC types 
are relatively more numerous at smaller galactocentric distances 
($R_G$, Massey \& Johnson 1998).  
Our selection should not be biased towards the selection of WC9 stars
as all WR subclasses show \ha\ , He{\sc ii} or C emission which falls
within our 70\ang\ bandpass (cf. discovery of WO star WR 93b, Drew et al 2004).
All 5 sharing the same WR subtype can therefore be seen as an indication of
the preponderance of late WC stars in the inner Milky Way.

However the fact remains that extrapolation inwards of the Solar Circle of the 
trend in the surface density of {\em all} types of WR star would indicate many 
more WR stars at $R_G < R_{\odot}$ might well be present than are actually 
known (see figure 10 in van der Hucht 2001)!  Accordingly, it is interesting 
and encouraging that this relatively shallow trawl of 4 SHS fields, based on 
red rather than either infrared or short wavelength optical data, has already 
produced 5 new strongly-reddened WR stars scattered across a mere $\sim$8 
square degrees, albeit embedded within a total search area of $\sim$48 square 
degrees. No two of the stars found are likely to be within the same cluster.
We conclude that optical searches of the Galactic Plane for new massive WR
stars remain worthwhile, if taken to greater depths of $R \sim 19$ that still 
lie well within the grasp of even 4-metre class telescopes.  

\paragraph*{ Acknowledgements}

EH and MP both acknowledge the support of postgraduate studentships funded by 
the Particle Physics \& Astronomy Research Council of the United Kingdom.
This paper makes use of: data obtained with the AAO/UK Schmidt Telescope at 
Siding Spring Observatory, NSW, Australia; the {\sc simbad} database, 
operated at CDS, Strasbourg, France; data products from the Two Micron All Sky 
Survey; data from the USNOFS Image and Catalogue archive; data from the Guide 
Star Catalogue II  and data from the DEep Near Infrared Survey (DENIS). We
also acknowledge the use of Starlink software in both the reduction and
analysis of our spectroscopic data.

\end{document}